\documentclass[11pt]{article}

\usepackage{amsmath}
\usepackage{theorem}
\usepackage{subfig}
\usepackage{epsfig}
\usepackage{amssymb}
\usepackage{verbatim}
\usepackage{fullpage}

\newtheorem{theorem}	 			{Theorem}

{\theorembodyfont{\rmfamily} \newtheorem{definition}
[theorem]	{Definition}}
{\theorembodyfont{\rmfamily} }
{\theorembodyfont{\rmfamily} }
{\theorembodyfont{\rmfamily} }
{\theorembodyfont{\rmfamily} }
\theoremstyle{break}
{\theorembodyfont{\rmfamily} }

\def\NN{\mathbb{N}}

\def\sse{\subseteq}

\title{Beyond Worst-Case Analysis}
\author{
Tim Roughgarden\thanks{Department of Computer Science, 
Stanford University, 474 Gates Building, 353 Serra Mall, Stanford, CA 94305.
Email: {\tt tim@cs.stanford.edu}.}}

\begin{document}

\maketitle


\section{Introduction}

Comparing different algorithms is hard.  For almost any pair of
algorithms and measure of algorithm performance---like running time or
solution quality---each algorithm will perform better
than the other on some inputs.\footnote{In rare cases a problem admits
  an {\em instance-optimal algorithm}, which is as good as every other
  algorithm on every input, up to a constant factor~\cite{FLN03}.  For most
  problems, there is no instance-optimal algorithm, and there is no
  escaping the incomparability of different algorithms.}
For example, the insertion sort
algorithm is faster than merge sort on already-sorted arrays
but slower on many other inputs.  When two algorithms have incomparable
performance, how can we deem one of them ``better than'' the other?

{\em Worst-case analysis} is a specific modeling choice in the
analysis of algorithms, where the overall performance of an algorithm
is summarized by its worst performance on any input of a given size.
The ``better'' algorithm is then the one with superior worst-case
performance.  Merge sort, with its worst-case asymptotic running
time of $\Theta(n \log n)$ for arrays of length $n$, is better in this
sense than insertion sort, which has a worst-case running time
of $\Theta(n^2)$.

While crude, worst-case analysis can be tremendously useful, and it is
the dominant paradigm for algorithm analysis in theoretical computer
science.  A good worst-case guarantee is the best-case scenario for an
algorithm, certifying its general-purpose utility and absolving its
users from understanding which inputs are relevant to their
applications.  Remarkably, for many fundamental computational
problems, there are algorithms with excellent worst-case performance
guarantees.  The lion's share of an undergraduate algorithms course 
comprises
algorithms that run in linear or near-linear time in the
worst case.

For many problems a bit beyond the scope of an undergraduate
course, however, the downside of worst-case analysis rears
its ugly head.  We next review three classical examples where
worst-case analysis gives misleading or useless advice about how to
solve a problem; further examples in modern machine learning are
described later.
These examples motivate the alternatives to
worst-case analysis described in the rest of the article.\footnote{For
  many more examples, analysis frameworks, and applications, see the
  author's lecture notes~\cite{bwca}.}  



\begin{figure}
\centering
\includegraphics[width=.35\textwidth]{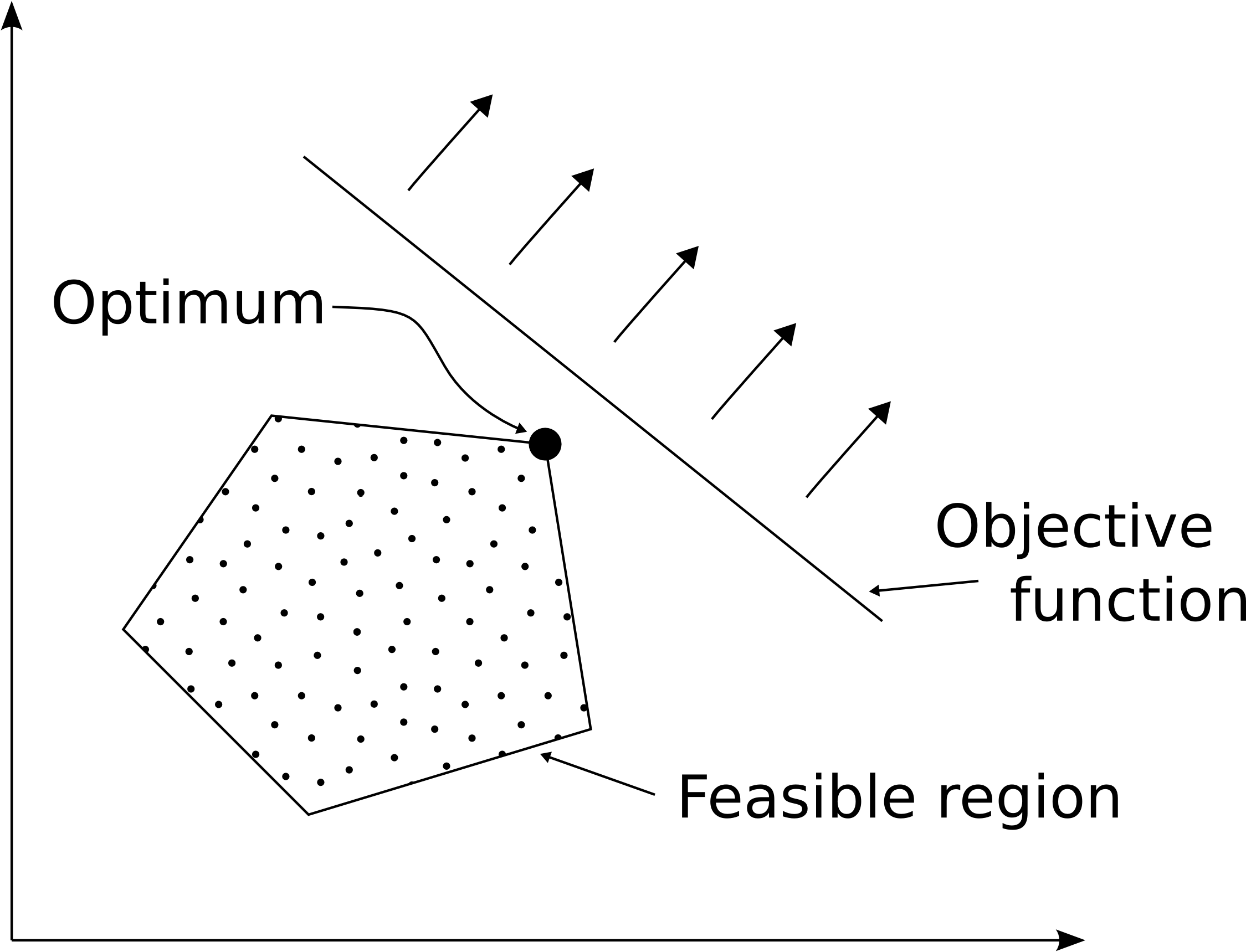}
\caption{A two-dimensional linear programming problem.}\label{f:lp}
\end{figure}

\paragraph{The simplex method for linear programming.}
Perhaps the most famous failure of worst-case analysis concerns linear
programming, the problem of optimizing a linear function subject to
linear constraints (Figure~\ref{f:lp}).  
Dantzig's {\em simplex method} is an algorithm from the 1940s
that solves linear programs using greedy local search
on the vertices on the solution set boundary, and
variants of it remain in wide use to this day.  The enduring appeal of
the simplex method stems from its consistently superb performance in
practice.  Its running time typically scales modestly with the input
size, and it routinely solves linear programs with millions of
decision variables and constraints.  This robust empirical performance
suggested that the simplex method might well solve every linear
program in a polynomial amount of time.

In 1972, Klee and Minty
showed by example that there are
contrived linear programs that force the simplex method to run in time
exponential in the number of decision variables (for all of the common
``pivot rules'' for choosing the next vertex).  This illustrates the
first potential pitfall of worst-case analysis: overly pessimistic
performance predictions that cannot be taken at face value.  The
running time of the simplex method is polynomial for all practical
purposes, despite the exponential prediction of worst-case analysis.

To add insult to injury, the first worst-case polynomial-time
algorithm for linear programming, the ellipsoid method,
is not competitive with the simplex method in
practice.\footnote{Interior-point methods,
developed five
  years later, lead to algorithms that both run in
  worst-case polynomial time and are competitive with the simplex
  method in practice.}
Taken at face value, worst-case analysis recommends the ellipsoid
method over the empirically superior simplex method.
One framework for narrowing the gap between these theoretical
predictions and empirical observations is {\em smoothed analysis},
discussed later in this article.

\begin{figure}
\centering
\includegraphics[width=.35\textwidth]{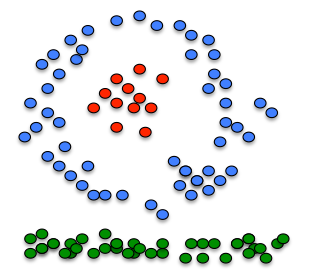}
\caption{One possible way to group data points into three clusters.}\label{f:clustering}
\end{figure}

\paragraph{Clustering and $NP$-hard optimization problems.}
Clustering is a form of unsupervised learning (finding patterns in
unlabeled data), where the informal goal is to partition a set of
points into ``coherent groups'' (Figure~\ref{f:clustering}).  
One popular way to coax this goal
into a well-defined computational problem is to posit a numerical
objective function over clusterings of the point set, and then seek
the clustering with the best objective function value.  For example,
the goal could be to choose $k$ cluster centers to minimize the sum of
the distances between points and their nearest centers (the $k$-median
objective) or the sum of the squared such distances (the $k$-means
objective).  Almost all natural optimization problems that are defined
over clusterings are $NP$-hard.

In practice, clustering is not viewed as a particularly
difficult problem.  Lightweight clustering algorithms, like Lloyd's
algorithm for $k$-means
and its variants,
regularly return the intuitively ``correct'' clusterings of real-world
point sets.  How can we reconcile the worst-case intractability of
clustering problems with the empirical success of relatively simple
algorithms?\footnote{More generally, optimization problems are more
  likely to be $NP$-hard than not.  In many cases, even computing an
  approximately optimal solution is an $NP$-hard problem (see
  Trevisan~\cite{T14}, for example).  Whenever an efficient algorithm
  for such a problem performs better on real-world instances than
  (worst-case) complexity theory would suggest, there's an opportunity
  for a refined and more accurate theoretical analysis.}

One possible explanation is that {\em clustering is hard only when it
  doesn't matter}~\cite{DLS12}.  For example, if the difficult
instances of an $NP$-hard clustering problem look like a bunch of
random unstructured points, who cares?  The common use case for a
clustering algorithm is for points that represent images, or
documents, or proteins, or some other objects where a ``meaningful
clustering'' is likely to exist.  Could instances with a meaningful
clustering be easier than worst-case instances?  This article surveys
recent theoretical developments that
support an affirmative answer.

\paragraph{Cache replacement policies.}  
Consider a system with a small fast memory (the cache) and a big slow
memory.  Data is organized into blocks called {\em pages}, with up to $k$
different pages fitting in the cache at once.  A page request results
in either a cache hit (if the page is already in the cache) or a cache
miss (if not).  On a cache miss, the requested page must be brought
into the cache.  If the cache is already full, then some page in it
must be evicted.  A cache policy is an algorithm for making these
eviction decisions.  Any systems textbook will recommend aspiring to
the least recently used (LRU) policy, which evicts the page whose most
recent reference is furthest in the past.  The same textbook will explain
why: real-world page request sequences tend to exhibit locality of
reference, meaning that recently requested pages are likely to be
requested again soon.  The LRU policy uses the recent past as a
prediction for the near future.  Empirically, it typically suffers
fewer cache misses than competing policies like first-in first-out
(FIFO).

Sleator and Tarjan~\cite{ST85} founded the 
area of {\em online algorithms}, which are algorithms that must
process their input as it arrives over time (like cache policies).
One of their first observations was that worst-case analysis,
straightforwardly applied, provides no useful insights about the
performance of different cache replacement policies.  For every
deterministic policy and cache size~$k$, there is a pathological page
request sequence that triggers a page fault rate of 100\%, even though
the optimal clairvoyant replacement policy (known as B\'el\'ady's
algorithm) would have a page fault rate of at most $(1/k)$\%.  This
observation is troublesome both for its absurdly pessimistic
performance prediction and for its failure to differentiate between
competing replacement policies (like LRU vs.\ FIFO).
One solution, described in the next section, is to choose an
appropriately fine-grained parameterization of the input space and to
assess and compare algorithms using parameterized guarantees.


\section{Models of Typical Instances}

Maybe we shouldn't be surprised that worst-case analysis fails to
advocate LRU over FIFO.  The empirical superiority of LRU is due to
the special structure in real-world page request sequences---locality
of reference---and traditional worst-case analysis provides no
vocabulary to speak about this structure.\footnote{If worst-case
  analysis has an implicit model of data, then it's the ``Murphy's
  Law'' data model, where the instance to be solved is an
  adversarially selected function of the chosen algorithm.  Outside of
  cryptographic applications, this is a rather paranoid and
  incoherent way to think about a computational problem.}  This is
what work on ``beyond worst-case analysis'' is all about:
\begin{itemize}
\item [] {\em articulating properties of ``real-world'' inputs, and proving
rigorous and meaningful algorithmic guarantees for inputs with these properties.}
\end{itemize}
Research in the area has both a scientific dimension, where
the goal is to develop transparent mathematical models that explain
empirically observed phenomena about algorithm performance, 
and an engineering dimension, where the goals are to provide accurate
guidance about which algorithm to use for a problem and to design new
algorithms that perform particularly well on the relevant inputs.

One exemplary result in beyond worst-case analysis is due to
Albers et al.~\cite{AFG02}, for the online paging problem described
in the introduction.  The key idea is to parameterize page request sequences
according to how much locality of reference they exhibit, and then
prove parameterized worst-case guarantees.  
Refining worst-case
analysis in this way leads to dramatically more informative
results.\footnote{Parameterized guarantees are common in the analysis
  of algorithms.  For example, the field of parameterized algorithms
  and complexity 
has developed a rich theory
  around parameterized running time bounds
(see the book by Cygan et al.~\cite{C+15}).
Theorem~\ref{t:main}
  employs an unusually fine-grained and problem-specific
  parameterization, and in exchange obtains unusually accurate and
  meaningful results.}

Locality of reference is quantified via the size of the working set of
a page request sequence.
Formally, for a function $f:\NN \rightarrow
\NN$, we say that a request
sequence {\em conforms to $f$} if, in every window of $w$ consecutive
page requests, at most $f(w)$ distinct pages are requested.  
For example, the identity function $f(w)=w$ imposes no restrictions on
the page request sequence.  A sequence can only conform to a sublinear
function like $f(w)=\lceil \sqrt{w} \rceil$ or $f(w) = \lceil 1 +
\log_2 w \rceil$ if it exhibits
locality of reference.\footnote{The notation $\lceil x \rceil$ means
  the number $x$, rounded up to the nearest integer.}

The following worst-case guarantee is parameterized by a number
$\alpha_f(k)$, between 0 and~1, that we discuss shortly; recall
that~$k$ denotes the cache size.
It assumes that the function~$f$ is ``concave'' in
the sense that the number of inputs with value~$x$ under~$f$
(that is, $|f^{-1}(x)|$) is nondecreasing in~$x$.
\begin{theorem}[Albers et al.~\cite{AFG02}]\label{t:main}
\mbox{}
\begin{itemize}

\item [(a)] For every $f$ and $k$
and every deterministic cache replacement policy, the worst-case page
  fault rate (over sequences that conform to $f$) is at least
  $\alpha_f(k)$.

\item [(b)] For every $f$ and $k$ and every sequence that conforms to $f$,
the page fault rate 
of the LRU policy is at most $\alpha_f(k)$.

\item [(c)] There exists a choice of $f$ and $k$, and a page request
  sequence that conforms to~$f$, such that the
page fault rate of the FIFO policy is strictly larger than $\alpha_f(k)$.

\end{itemize}
\end{theorem}
Parts~(a) and~(b) prove the worst-case optimality of the LRU policy
in a strong sense, $f$-by-$f$ and $k$-by-$k$.
Part~(c) differentiates LRU from FIFO, as the latter is suboptimal for
some (in fact, many) choices of $f$ and $k$.

The guarantees in Theorem~\ref{t:main} are so
good that they are meaningful even when taken at face value---for
sublinear~$f$'s, $\alpha_f(k)$ goes to~0 reasonably quickly with~$k$.
For example, if $f(w) = \lceil \sqrt{w} \rceil$, then $\alpha_f(k)$ scales
with $1/\sqrt{k}$.
Thus with a cache size of 10,000, the page fault
rate is always at most 1\%.
If $f(w) = \lceil 1 + \log_2 w \rceil$, then $\alpha_f(k)$ goes to~0
even faster with~$k$, roughly as $k/2^k$.\footnote{See Albers et
  al.~\cite{AFG02} for the precise closed-form formula for
  $\alpha_f(k)$ in general.}

\section{Stable Instances}\label{s:stable}


Are point sets with meaningful clusterings easier to cluster than
worst-case point sets?
We next describe one way to define a ``meaningful clustering,'' due
to Bilu and Linial~\cite{BL10}; for others,
see Ackerman and Ben{-}David~\cite{AB09},
Balcan et al.~\cite{BBG13}, Daniely et al.~\cite{DLS12}, 
Kumar and Kannan~\cite{KK10}, and
Ostrovsky et al.~\cite{O+06}.


\paragraph{The maximum cut problem.}
Suppose you have a bunch of data points representing images of cats
and images of dogs, and you'd like to automatically discover these two
groups.  One approach is to reduce this task to the {\em maximum cut}
problem, where the goal is to partition the vertices~$V$ of a
graph~$G$ with edges~$E$ and nonnegative edge weights into two
groups, while maximizing the total weight of the edges that have one
endpoint in each group.  The
reduction forms a complete graph~$G$, with vertices corresponding to
the data points, and assigns a weight~$w_e$ to each edge~$e$
indicating how dissimilar its endpoints are.  The maximum cut of~$G$
is a 2-clustering that tends to put dissimilar pairs of points in
different clusters.

\begin{figure}
\centering
\subfloat[The maximum cut]{\includegraphics[scale=.45]{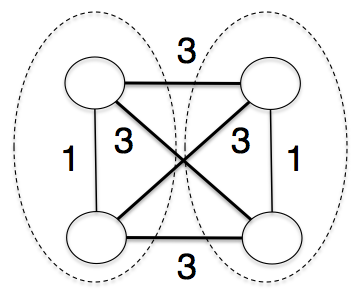}}
\qquad
\subfloat[Still the maximum cut]{\includegraphics[scale=.45]{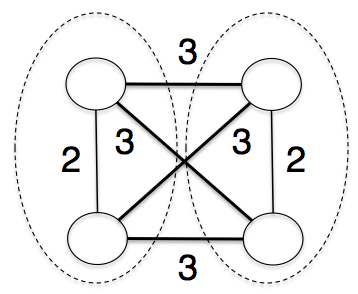}}
\caption{In a perturbation-stable maximum cut instance, the optimal
  solution is   invariant under small perturbations to the edges' weights.}
\label{f:maxcut}
\end{figure}

There are many ways to quantify ``dissimilarity'' between images, and
different definitions might give different optimal 2-clusterings of
the data points.  One would hope that, for a range of reasonable
measures of dissimilarity, the maximum cut in the example above would
have all cats on one side and all dogs on the other.  In other words,
the maximum cut should be invariant under minor changes to the
specification of the edge weights (Figure~\ref{f:maxcut}).
\begin{definition}[Bilu and Linial~\cite{BL10}]\label{d:ps}
An instance~$G=(V,E,w)$ of the maximum cut problem is {\em
  $\gamma$-perturbation stable} if, for all ways of multiplying the
weight $w_e$ of each edge $e$ by a factor $\alpha_e \in [1,\gamma]$,
the optimal solution remains the same.
\end{definition}
A perturbation-stable instance has a ``clearly optimal''
solution---a uniqueness assumption on steroids---thus
formalizing the idea of a ``meaningful clustering.''  In machine
learning parlance, perturbation stability can be viewed as a type of
``large margin'' assumption.

The maximum cut problem is $NP$-hard in general.
But what about the special case of $\gamma$-perturbation-stable instances?
As $\gamma$ increases, fewer and fewer instances qualify as
$\gamma$-perturbation stable.  Is there a sharp stability
threshold---a value of $\gamma$ where the maximum cut problem switches
from $NP$-hard to polynomial-time solvable?

Makarychev et al.~\cite{MMV14} largely resolved this question.  On the
positive side, they showed that if~$\gamma$ is at least a slowly
growing function of the number of vertices~$n$, then the maximum cut
problem can be solved in polynomial time for all $\gamma$-perturbation
stable instances.\footnote{Specifically,
  $\gamma = \Omega(\sqrt{\log n} \log \log n)$.}  Makarychev et
al.~\cite{MMV14} use
techniques from the field of metric embeddings to show that, in such
instances, the unique optimal solution of a certain semidefinite
programming relaxation corresponds precisely to the maximum
cut.\footnote{In general, the optimal solution of a linear or semidefinite
  programming relaxation of an $NP$-hard problem is a ``fractional
  solution'' that does not correspond to a feasible solution to the
  original problem.}  Semidefinite programs are convex programs, and
can be solved to arbitrary precision in polynomial time.  There is
also evidence that the maximum cut cannot be recovered in polynomial
time in $\gamma$-perturbation-stable instances for much smaller values
of~$\gamma$~\cite{MMV14}.

\paragraph{Other clustering problems.}
Bilu and Linial~\cite{BL10} defined $\gamma$-perturbation-stable
instances specifically for the maximum cut problem, but the definition
makes sense more generally for any optimization problem with a linear
objective function.  The study of $\gamma$-perturbation-stable
instances has been particularly fruitful for $NP$-hard clustering
problems in metric spaces, where interpoint distances are required to
satisfy the triangle inequality.  Many such problems, including the
$k$-means, $k$-median, and $k$-center problems, are polynomial-time
solvable already in 2-perturbation-stable
instances~\cite{AMM17,BHW16}.  The algorithm in Angelidakis et
al.~\cite{AMM17}, like its precursor in Awasthi et al.~\cite{ABS11},
is inspired by the widely-used single-linkage clustering algorithm.
It computes a minimum spanning tree (where edge weights are the
interpoint distances) and uses dynamic programming to optimally remove
$k-1$ edges to define $k$ clusters.  To the extent that we're
comfortable identifying ``instances with a meaningful clustering''
with 2-perturbation-stable instances, these results give a precise
sense in which clustering is hard only when it doesn't
matter.\footnote{A relaxed and more realistic version of
  perturbation-stability allows small perturbations to make small
  changes to the optimal solution.
Many of the results mentioned
  in this section can be extended to instances meeting this
  relaxed condition, with a polynomial-time algorithm guaranteed to
  recover a solution that closely resembles the optimal
  one~\cite{AMM17,BBG13,MMV14}.}

\paragraph{Overcoming $NP$-hardness.}
Polynomial-time algorithms for $\gamma$-perturbation-stable instances
continue the age-old tradition of identifying ``islands of
tractability,'' meaning polynomial-time solvable special cases of
$NP$-hard problems.  Two aspects of these results diverge
from a majority of 20th-century research on tractable special cases.
First, perturbation-stability is not an easy condition to check, in
contrast to a restriction like graph planarity or
Horn-satisfiability.  Instead, the assumption is justified with a
plausible narrative about why ``real-world instances'' might satisfy
it, at least approximately.  Second, in most work going beyond
worst-case analysis, the goal is to study general-purpose algorithms,
which are well defined on all inputs, and use the assumed instance
structure only in the algorithm analysis (and not explicitly in its design).  
The hope is that the algorithm continues to perform well on many
instances not covered by its formal guarantee.  
The results
above for mathematical programming relaxations and
single-linkage-based algorithms are good examples of this paradigm.

\paragraph{Analogy with sparse recovery.}
There are compelling parallels between the recent research on
clustering in stable instances and slightly older results in a field
of applied mathematics known as {\em sparse recovery}, where the goal
is to reverse engineer a ``sparse'' object from a small number of
clues about it.
%
A common theme in both areas is identifying relatively weak conditions
under which a tractable mathematical programming relaxation of an
$NP$-hard problem is guaranteed to be exact, meaning that the original
problem and its relaxation have the same optimal solution.

For example, a canonical problem in sparse recovery is {\em
  compressive sensing}, where the goal is to recover an unknown sparse
signal (a vector of length~$n$) from a small number~$m$ of linear
measurements of it.  Equivalently, given an
$m \times n$ measurement matrix $A$ with $m \ll n$ and the measurement
results $b = Az$, the problem is to figure out 
the signal~$z$.  This problem has
several important applications, for example in medical imaging.
If $z$ can be arbitrary, then the problem is hopeless: since $m < n$,
the linear system $Ax = b$ is underdetermined and has an infinite
number of solutions (of which $z$ is only one).  But many real-world
signals are (approximately) $k$-sparse in a suitable basis for small $k$,
meaning that (almost) all of the mass
is concentrated on $k$ coordinates.\footnote{For example, audio
  signals are typically approximately sparse in the Fourier basis,
  images in the wavelet basis.}  The main results in compressive
sensing show that, under appropriate assumptions on~$A$, 
the problem can be solved efficiently even when $m$ is only
modestly bigger than $k$ (and much smaller than $n$)~\cite{CRT06,D06}.
One way to prove these results is to formulate a linear programming
relaxation 
of the ($NP$-hard) problem
of computing the sparsest solution to $Ax=b$,
and then show that this relaxation is exact.


\section{Planted and Semi-Random Models}\label{s:planted}


Our next genre of models
is also inspired by the idea that
interesting instances of a problem should have ``clearly optimal''
solutions, but differs from the above stability conditions
in assuming a generative model---a specific distribution over inputs.
The goal is to design an algorithm that, with high probability
over the assumed input distribution, computes an optimal solution in
polynomial time.

\paragraph{The planted clique problem.}
In the {\em maximum clique} problem, the input is an undirected graph
$G=(V,E)$, and the goal is to identify the largest subset of vertices
that are mutually adjacent.  This problem is $NP$-hard, even to
approximate by any reasonable factor.  Is it easy when there is a
particularly prominent clique to be found?

Jerrum~\cite{J92} suggested the following generative model.  There is
a fixed set $V$ of $n$ vertices.  First, each possible edge $(u,v)$ is
included independently with 50\% probability.  This is also known as
an Erd\"os-Renyi random graph with edge density~$\tfrac{1}{2}$.
Second, for a parameter $k \in \{1,2,\ldots,n\}$, a subset $Q \sse V$
of $k$ vertices is chosen uniformly at random, and all remaining edges
with both endpoints in $Q$ are added to the graph (thus making $Q$ a
$k$-clique).

How big does $k$ need to be before $Q$ becomes visible to a
polynomial-time algorithm?  The state-of-the-art is a spectral
algorithm of Alon et al.~\cite{AKS98}, which
recovers the planted clique~$Q$ with high probability provided $k$ is
at least a constant times $\sqrt{n}$.  
Recent work suggests that efficient algorithms cannot recover~$Q$
for significantly
smaller values of~$k$~\cite{sos}.

\begin{figure}
\begin{center}
\includegraphics{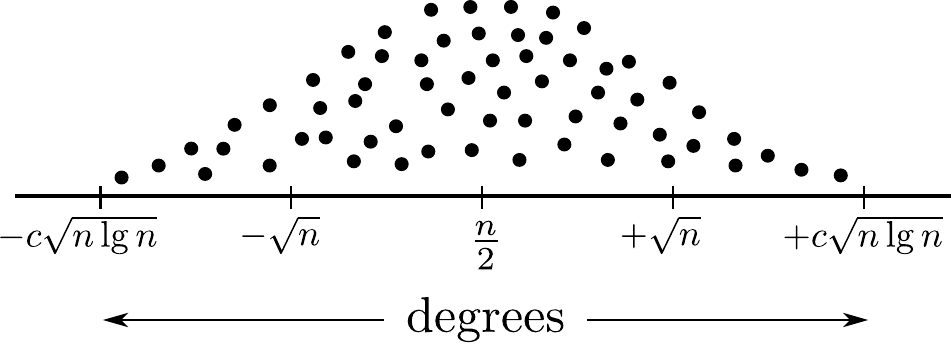}
\caption{Degree distribution of an Erd\"os-Renyi random
graph with edge density $\tfrac{1}{2}$,
before planting
the $k$-clique clique~$Q$. 
If $k = \Omega(\sqrt{n \lg n})$, then the planted clique will consist
  of the $k$ vertices with the highest degrees.}\label{f:catapult} 
\end{center}
\end{figure}

\paragraph{An unsatisfying algorithm.}  The algorithm of Alon et
al.~\cite{AKS98} is theoretically interesting and plausibly useful.
But if we take $k$ to be just a bit bigger, at least a constant times
$\sqrt{n \log n}$, then there is an uninteresting and useless
algorithm that recovers the planted clique with high probability:
return the $k$ vertices with the largest degrees.  To see why this
algorithm works, think first about the sampled Erd\"os-Renyi random
graph, before the clique~$Q$ is planted.  The expected degree of each
vertex is $\approx n/2$, with standard deviation $\approx \sqrt{n}/2$.
Textbook large deviation inequalities show that, with high
probability, the degree of every vertex is within
$\approx \sqrt{\ln n}$ standard deviations of its expectation
(Figure~\ref{f:catapult}).  Planting a clique~$Q$ of size
$a \sqrt{n \log n}$, for a sufficiently large constant~$a$, then
boosts the degrees of all of the clique vertices enough that they
catapult past the degrees of all of the non-clique vertices.

What went wrong?  The same thing that often goes wrong with pure
average-case analysis---the solution is brittle and overly tailored to
a specific distributional assumption.  How can we change the input
model to encourage the design of algorithms with more robust
guarantees?  Can we find a sweet spot between average-case and
worst-case analysis?

\paragraph{Semi-random models.}
Blum and Spencer~\cite{blum-spencer} proposed studying {\em
  semi-random} models, where nature and an adversary collaborate to
produce an input.  In many such models, nature first samples an input
from a specific distribution (like the probabilistic planted clique
model above), which is then modified by the adversary before being
presented as an input to an algorithm.  It is important to restrict
the adversary's power, so that it cannot simply throw out
nature's starting point and replace it with a worst-case instance.
Feige and Killian~\cite{FK} suggested studying {\em monotone
  adversaries}, which can only modify the input by making the optimal
solution ``more obviously optimal.''  For example, in the semi-random
version of the planted clique problem, a monotone adversary is only
allowed to remove edges that are not in the planted clique~$Q$---it
cannot remove edges from~$Q$ or add edges outside~$Q$.

Semi-random models with a monotone adversary may initially seem no
harder than the planted models that they generalize.  But 
let's return to the planted clique model with
$k = \Omega(\sqrt{n \log n})$, where the
``top-$k$ degrees'' algorithm succeeds with high probability when
there is no adversary.  A
monotone adversary can easily foil this algorithm in the semi-random
planted clique model,
by removing edges between clique and non-clique vertices to decrease
the degrees of the former back down to $\approx n/2$.  Thus the
semi-random model forces us to develop smarter, more robust
algorithms.\footnote{The extensively studied ``stochastic block
  model''generalizes the planted clique model (see
  e.g.~Moore~\cite{M17}), and is another fruitful playground for
  semi-random models.  Here, the vertices of a graph are partitioned
  into groups, and the probability that an edge is present is a
  function of the groups that contain its endpoints.  The
  responsibility of an algorithm in this model is to recover the
  (unknown) vertex partition.  This goal becomes provably strictly
  harder in the presence of a monotone adversary~\cite{MPW16}.}

For the semi-random planted clique model, Feige and
Krauthgamer~\cite{FK} gave a polynomial-time algorithm that recovers
the clique with high probability provided $k=\Omega(\sqrt{n})$.  The
spectral algorithm by Alon et al.~\cite{AKS98} achieved this guarantee
only in the standard planted clique model, and it does not provide any
strong guarantees for the semi-random model.  The algorithm of Feige
and Krauthgamer~\cite{FK} instead uses a semidefinite programming
relaxation of the problem.
Their analysis shows that this relaxation
is exact with high probability in the standard planted clique model
(provided $k = \Omega(\sqrt{n})$), and uses the monotonicity
properties of optimal mathematical programming solutions to argue that
this exactness cannot be sabotaged by any monotone adversary.


\section{Smoothed Analysis}\label{s:sa}

{\em Smoothed analysis} is another example of a semi-random model, now
with the order of operations reversed: an adversary goes first and
chooses an arbitrary input, which is then perturbed slightly by nature.
Smoothed analysis can be applied to any problem where ``small
perturbations'' make sense, including most problems with real-valued
inputs.  It can be applied to any measure of algorithm performance,
but has proven most effective for running time
analyses.

Like other semi-random models, smoothed analysis has the
benefit of potentially escaping worst-case inputs (especially if they
are ``isolated''), while avoiding overfitting a solution to a specific
distributional assumption.  There is also a plausible narrative about
why ``real-world'' inputs are captured by this framework: whatever
problem you'd like to solve, there are inevitable inaccuracies in its
formulation (from measurement error, uncertainty, and so on).

\paragraph{The simplex method.}
Spielman and Teng~\cite{ST04} developed the smoothed analysis
framework with the specific goal of proving that bad inputs for the
simplex method are exceedingly rare.  
Average-case analyses of the simplex method from the 1980s
(e.g., Borgwardt~\cite{borgwardt})
provide evidence for this thesis, but smoothed analysis provides 
more robust support for it.

The perturbation model in Spielman and Teng~\cite{ST04} is:
independently for each entry of the constraint matrix and right-hand
side of the linear program, add a Gaussian (i.e., normal) random
variable with mean~0 and standard deviation $\sigma$.\footnote{This
  perturbation results in a dense constraint matrix even if the
  original one was sparse, and for this reason Theorem~\ref{t:st04} is
  not fully satisfactory.  Extending this result to
sparsity-preserving perturbations is an
  important open question.}  The parameter~$\sigma$ interpolates between
worst-case analysis (when $\sigma=0$) and pure average-case analysis
(as $\sigma \rightarrow \infty$, the perturbation drowns out the
original linear program).  The main result states that the expected
running time of the simplex method is polynomial as long as typical
perturbations have magnitude at least an inverse polynomial function
of the input size (which is small!).
\begin{theorem}[Spielman and Teng~\cite{ST04}]\label{t:st04}
  For every initial linear program, in expectation over the
  perturbation to the program, the running time of the simplex method
  is polynomial in the input size and in $\tfrac{1}{\sigma}$.
\end{theorem}
The running time blow-up as $\sigma \rightarrow 0$ is necessary
because the worst-case running time of the simplex method is
exponential.  Several researchers have devised
simpler analyses and better polynomial running times, most recently
Dadush and Huiberts~\cite{DH18}.
All of these analyses are for a specific pivot rule, the ``shadow
pivot rule.''  The idea is to project the high-dimensional feasible
region of a linear program onto a plane (the ``shadow'') and run the
simplex method there.  The hard part of proving Theorem~\ref{t:st04} is
showing that, with high probability over nature's perturbations, the
perturbed instance is ``well-conditioned'' in the sense that each step
of the simplex method makes significant progress traversing the
boundary of the shadow.

\paragraph{Local search.}
A local search algorithm for an optimization problem maintains a
feasible solution, and iteratively improves that solution via ``local
moves'' for as long as possible, terminating with a locally optimal
solution.  Local search heuristics are ubiquitous in practice, in many
different application domains.  Many such heuristics have an
exponential worst-case running time, despite always terminating
quickly in practice (typically within a sub-quadratic number of
iterations).  Resolving this disparity is right in the wheelhouse of
smoothed analysis.  For example, Lloyd's algorithm for the $k$-means
problem can require an exponential number of iterations to converge in
the worst case, but needs only an expected polynomial number of
iterations in the smoothed case (see~Arthur et al.~\cite{AMR11} and
the references therein).\footnote{An orthogonal issue with local
  search heuristics is the possibility of outputting a locally optimal
  solution that is much worse than a globally optimal one.  Here, the
  gap between theory and practice is not as embarrassing---for many
  problems, local search algorithms really can produce pretty lousy
  solutions.  For this reason, one generally invokes a local search
  algorithm many times with different starting points and returns the
  best of all of the locally optimal solutions found.}

Much remains to be done, however.
For a concrete challenge
problem, let's revisit the maximum cut problem.
The input is an undirected graph $G=(V,E)$ with edge weights, and the
goal is to partition~$V$ into two groups to maximize the total weight
of the edges with one endpoint in each group.  Consider a local search
algorithm that modifies the current solution by moving a single vertex
from one side to the other (known as the ``flip neighborhood''), and
performs such moves as long as they increase the sum of the weights of
the edges crossing the cut.  In the worst case, this local search
algorithm can require an exponential number of iterations to converge.
What about in the smoothed analysis model,
where a small random perturbation is added to each edge's weight?  The
natural conjecture is that local search should terminate in a polynomial
number of iterations, with high probability over the perturbation.
This conjecture has been proved
for graphs with maximum degree $O(\log n)$~\cite{ET11} and for the
complete graph~\cite{fan}; for general graphs, the state-of-the-art is
a quasi-polynomial-time guarantee (meaning $n^{O(\log n)}$
iterations)~\cite{ER17}.

More ambitiously, it is tempting to speculate that for {\em every}
natural local search problem, local search terminates in 
a polynomial number of iterations in the smoothed analysis model (with
high probability).  
Such a result would be a huge success story for smoothed analysis and
beyond worst-case analysis more generally.

\section{On Machine Learning}\label{s:ml}

Much of the present and future of research going beyond worst-case
analysis is motivated by advances in machine
learning.\footnote{Arguably, even the overarching goal of research in
  beyond worst-case analysis---%
determining the best algorithm for an application-specific special
  case of a problem---is fundamentally a machine learning
  problem~\cite{GR16}.}
The unreasonable effectiveness of modern machine learning algorithms
has thrown the gauntlet down to algorithms researchers, and there is
perhaps no other problem domain with a more urgent need for the
“beyond worst-case” approach.

To illustrate some of the challenges, consider a canonical supervised
learning problem, where a learning algorithm is given a data set of
object-label pairs and the goal is to produce a classifier that
accurately predicts the label of as-yet-unseen objects (e.g., whether
or not an image contains a cat).  Over the past decade, aided by
massive data sets and computational power,
deep neural networks have achieved impressive levels of performance
across a range of prediction tasks~\cite{goodfellow}.  Their
empirical success flies in the face of conventional wisdom in multiple
ways.  First, most neural network training algorithms use first-order
methods (i.e., variants of gradient descent) to solve nonconvex
optimization problems that had been written off as computationally
intractable.  Why do these algorithms so often converge quickly to a
local optimum, or even to a global optimum?\footnote{See Jin et
  al.~\cite{J+17} and the references therein for recent progress on
  this question.}  Second, modern neural networks are typically
over-parameterized, meaning that the number of free parameters
(weights and biases) is considerably larger than the size of the
training data set.  Over-parameterized models are vulnerable to
large generalization error
(i.e., overfitting), but state-of-the-art neural networks generalize
shockingly well~\cite{Z+17}.  How can we explain this?  The answer
likely hinges on special properties of both real-world data sets and
the optimization algorithms used for neural network training
(principally
stochastic gradient descent).\footnote{See Neyshabur~\cite{N17} and
  the references therein for the latest developments in this direction.}

Another interesting case study, this time in unsupervised learning, 
concerns topic modeling.  The goal here is to process a
large unlabeled corpus of documents and produce a list of meaningful
topics and an assignment of each document to a mixture of topics.  One
computationally efficient approach to the problem is to use a singular
value decomposition subroutine to factor the term-document matrix into
two matrices, one that describes which words belong to which topics,
and one indicating the topic mixture of each document~\cite{P+00}.
This approach can lead to negative entries in the matrix factors,
which hinders interpretability.
Restricting the matrix factors to be nonnegative yields a problem that
is $NP$-hard in the worst case, but Arora et al.~\cite{A+13} gave a
practical factorization algorithm for topic modeling that
runs in polynomial time under a reasonable assumption about the data.
Their assumption states that each topic has at least one ``anchor
word,'' the presence of which strongly indicates that the document is
at least partly about that topic (such as the word ``Durant'' for the
topic ``basketball'').  Formally articulating this property of data
was an essential step in the development of their algorithm.

The beyond worst-case viewpoint can also contribute to machine
learning by ``stress-testing'' the existing theory and providing a
road map for more robust guarantees.  While work in beyond worst-case
analysis makes strong assumptions relative to the norm in theoretical
computer science, these assumptions are usually weaker than the norm
in statistical machine learning.  Research in the latter field often
resembles average-case analysis, for example when data points are
modeled as independent and identically distributed samples from some
(possibly parametric) distribution.  
The semi-random models described earlier in this article are
role models in blending adversarial and average-case modeling to
encourage the design of algorithms with robustly good performance.
Recent progress in computationally efficient robust statistics shares
much of the same spirit~\cite{D+17}.

\section{Conclusions}

With algorithms, silver bullets are few and far between.  No one
design technique leads to good algorithms for all computational
problems.  Nor is any single analysis framework---worst-case analysis
or otherwise---suitable for all occasions.  A typical algorithms
course teaches several 
paradigms for algorithm {\em design}, along with guidance about when
to use each of them; the field of beyond worst-case analysis holds the
promise of a comparably diverse toolbox for algorithm {\em analysis}.

Even at the level of a specific problem, there is generally no
magical, always-optimal algorithm---the best algorithm for the job
depends on the instances of the problem most relevant to the specific
application.  Research in beyond worst-case analysis acknowledges this
fact while retaining the emphasis on robust guarantees that is central
to worst-case analysis.  The goal of work in this area is to develop
novel methods for articulating the relevant instances of a problem,
thereby enabling rigorous explanations of the empirical performance of
known algorithms, and also guiding the design of new algorithms
optimized for the instances that matter.

With algorithms increasingly dominating our world, the need to
understand when and why they work has never been greater.  
The field of beyond worst-case analysis has already produced 
several striking results, but there remain many unexplained gaps
between the theoretical and empirical performance of widely-used
algorithms.  With so many opportunities for consequential research,
I suspect that the best work in the area is yet to come.


\section{Acknowledgments}

I thank Sanjeev Arora, Ankur Moitra, Aravindan Vijayaraghavan, and
four anonymous reviewers for several helpful suggestions.  This work
was supported in part by NSF award CCF-1524062, a Google Faculty
Research Award, and a Guggenheim Fellowship.

{\small


}

\end{document}